\documentclass[twocolumn]{aa}
\usepackage{graphicx}
\usepackage{txfonts}
\begin{document}
   \title{Discovery of a stellar system in the background of 47
   Tucanae\thanks{Based on data collected at the European Southern
   Observatory, La Silla, Chile, with the Wide Field Imager under the
   observing programmes 62.L-0354 and 64.L-0439. Also based on Wide
   Field Imager data retrieved from the ESO/ST-ECF Science Archive
   Facility.}}

   \subtitle{A new cluster of the Small Magellanic Cloud?}

   \author{M. Bellazzini
          \inst{1},
	  E. Pancino
	  \inst{1}
          \and
	  F.R. Ferraro
	  \inst{2}
          }
   \authorrunning{Bellazzini et al.}

   \offprints{M. Bellazzini}

   \institute{INAF - Osservatorio Astronomico di Bologna
              Via Ranzani 1, I-40127, Bologna, Italy\\
              \email{michele.bellazzini, elena.pancino@bo.astro.it}
         \and
             Universit\`a di Bologna, Dipartimento di Astronomia,
	     Via Ranzani 1, I-40127, Bologna, Italy\\
             \email{francesco.ferraro3@unibo.it}
             }

   \date{Received ; accepted}

   \abstract{We report on the discovery of a stellar system in the
   background of the Galactic globular cluster 47 Tucanae (NGC 104),
   located $14.8 \arcmin$ North-West of the cluster center. The object,
   whose apparent diameter is $D\simeq 30\arcsec$, is partially resolved
   into stars on the available CCD images, reaching a limiting magnitude
   of $V\sim 22.5$, and is detected as a  significant (more than $5
   \sigma$) overdensity of blue stars ($B-V<0.7$). The color magnitude
   diagram of the system, its characteristic projected size and its
   position in the sky suggest that it is an intermediate-old age
   cluster belonging to the Small Magellanic Cloud, whose outskirts lie
   in the background of 47 Tuc. Although less likely, the
   possibility that the object is an unknown dwarf galaxy in the
   outskirts of the Local Group cannot completely be ruled out by the
   present data.   

   }

   \maketitle
%

\section{Introduction}

The advent of modern surveys exploring large swaths of the sky at
different wavelenghts, that provided deeper insight into poorly
explored regions of the Galaxy and its surroundings, has lead to a
burst of discoveries and cataloguing of new stellar systems and
structures that has no precedent in the past three decades. For
example, the insight into high extinction regions provided by infrared
imaging (2MASS, Spitzer) has allowed the recognition of several stellar
systems that were hidden into the Galactic disc, or in general, in low
galactic latitude fields (see, for example \cite{bd3a,bd3b,glimpse}).

Also, well known high-surface-brightness objects subtending large solid
angles in the sky, as some nearby globular clusters or for instance
M31, may hide behind their body resolved stellar systems as star
clusters or dwarf galaxies (see \cite{whi}). Here we briefly report on
the curious case of a stellar system we discovered in the background of
the nearby globular cluster 47 Tucanae.

\section{Observations}

Observations were obtained at the 2.2 m ESO-MPI telescope at ESO (La
Silla) in July 1999 with the Wide Field Imager (WFI) and an additional
dataset was also retrieved from the ESO/ST-ECF Science Archive. The
final set consisted of several B, V and I images (six or seven 60s
exposures in each filter) centered on the globular cluster 47~Tucanae. 
Details of the data reduction procedure and other results obtained from
this dataset can be found in \cite{tucbss} and \cite{tip2}.

\begin{figure}
   \centering
   \includegraphics[width=5.5cm]{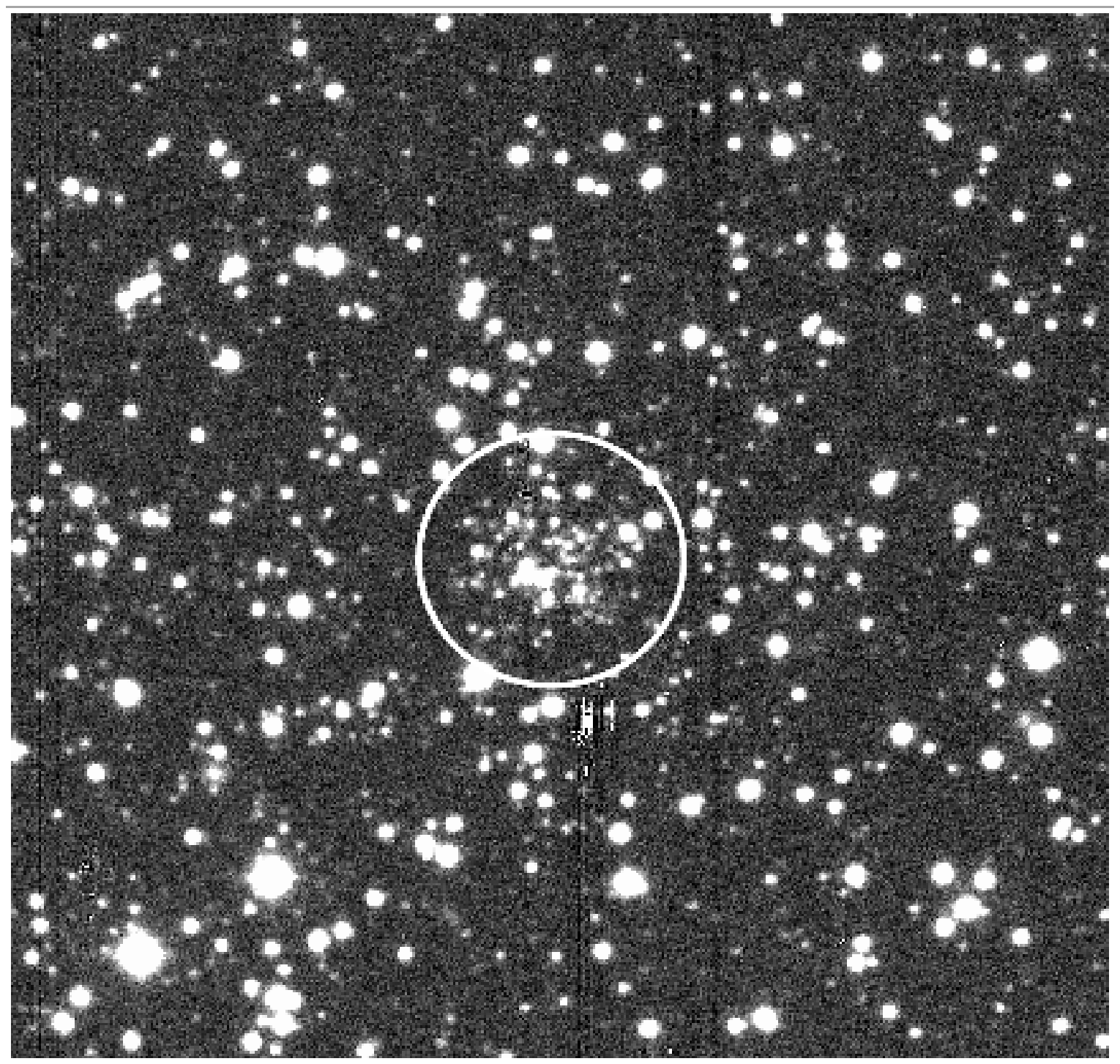}
   \includegraphics[width=5.5cm]{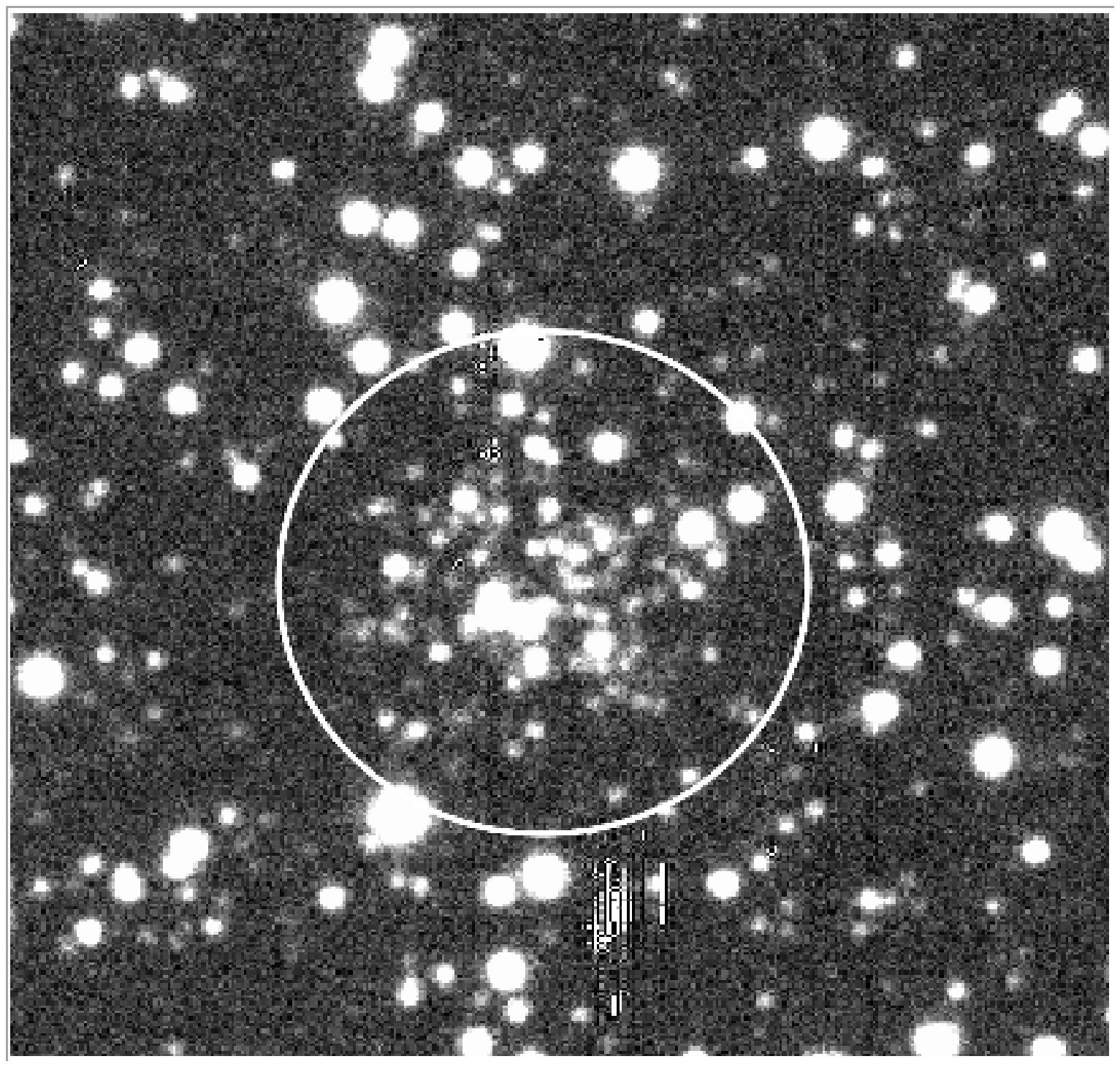}
      \caption{Upper panel: Combined B image of $2.7\arcmin \times
      2.7\arcmin$  centered on the newly discovered stellar system.
      North is up and East is left. Lower panel: The same region zoomed
      in by a factor of $\sim 2$. The overplotted circles have a
      diameter of $40\arcsec$. Virtually all the bright stars present
      in the image are members of the globular cluster 47 Tucanae.}
         \label{Fig1}
   \end{figure}

The WFI is a mosaic of eight $2048\times 4096$ pixel chips with a scale
of $0.238\arcsec$ pixel$^{-1}$. The whole mosaic covers an area of
$\simeq 34\arcmin \times 33\arcmin$. During the visual inspection of
the images we noted, on chip 4 of the B and V frames, an unusual bright
spot. A closer look revealed the presence of an anomalous clustering of
faint stars, placed over a barely visible concentration of unresolved
light. By averaging the three best seeing B images (chip 4 only), we
obtained the combined image shown in Fig.~\ref{Fig1}, revealing that we
have found a partially resolved stellar system, in the background of
the much more nearby cluster 47~Tuc, which for brevity we named
Bologna~A (Bo~A). 

%

\section{The nature of Bo~A}

As can be noticed from Fig.~\ref{Fig1}, the overall shape of Bo~A is
roundish and the apparent diameter is $\sim 30\arcsec$.  The position
of the optical center of simmetry is $RA = 0^h 21^m 30.5^s$ and $Dec =
-71\degr 56\arcmin 7\arcsec$ (equinox 2000.0). We searched the NED,
SIMBAD and Vizier databases as well as other databases (as for instance
the catalogue of SMC clusters by \cite{bd00}, hereafter BD00) and we
found no known object within $2 \arcmin$ of this position, except for a
few bright stars belonging to 47~Tuc. Bo~A is clearly visible in all of
our B and V images but is hard to recognise in I images, suggesting
that it is dominated by blue stars. It is also (barely) visible in the
DSS-2 images.

In the following we discuss the properties and nature of Bo~A as can be
deduced from our dataset.

   \begin{figure}
   \centering
   \includegraphics[width=8cm]{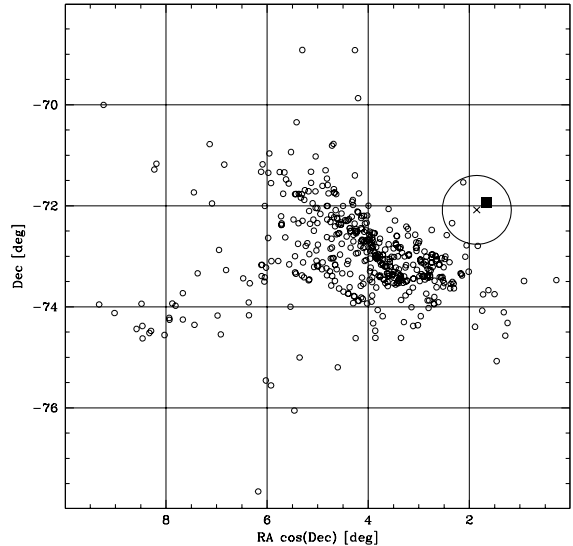}
   \caption{Projected map of all the confirmed SMC clusters from the
   catalogue by \cite{bd00} (open circles). The cross marks the
   position of the center of 47~Tuc, the circle has the radius equal to
   its tidal radius. Bo~A is represented by the filled square.}
              \label{Fig5}%
    \end{figure}
%
%

\subsection{Position in the sky}

It is well known that 47~Tuc is projected onto the outskirts of the
Small Magellanic Cloud (SMC), a galaxy which has a rich cluster system 
(469 confirmed members, according to BD00). Fig.~\ref{Fig5} shows the
sky-projected map of the confirmed SMC clusters from the catalogue of
BD00 (open circles). The center (\cite{dm}) and tidal radius
(\cite{tdk}) of 47~Tuc are marked by a cross and a large circle,
respectively. Bo~A is represented by the filled square: clearly, its
position is fully compatible with that of a SMC cluster. Hence, the
most natural hypothesis that can indeed be made is that Bo~A is a
previously unknown SMC cluster, projected by chance near 47~Tuc. The
postition of Bo~A can explain why it has never been identified before,
lying only just $\simeq 14.8\arcmin$ away from the center of this
cluster. The previous record of proximity to 47~Tuc was hold by HW5
(\cite{hw74}), located $34.6\arcmin$ away from the center.

\subsection{Color Magnitude Diagram}

Since the previous published photometric analysis of this dataset was
aimed at obtaining the cleanest possibile sample of stars in 47~Tuc
(\cite{tip2,tucbss}), only stars that were detected in at least three
images per passband were included in the final catalogues. Here we
noted that the stellar population of Bo~A appears faint and blue,
therefore it is not clearly visible in the shallow exposures, and it is
barely visible on the I band images. Therefore, we took the best seeing
B and V images of chip 4 ($FWHM=0.8\arcsec$) and performed a new
photometry with less restrictive requirements, to reach as deep as
possible.

   \begin{figure}
   \centering
   \includegraphics[width=8cm]{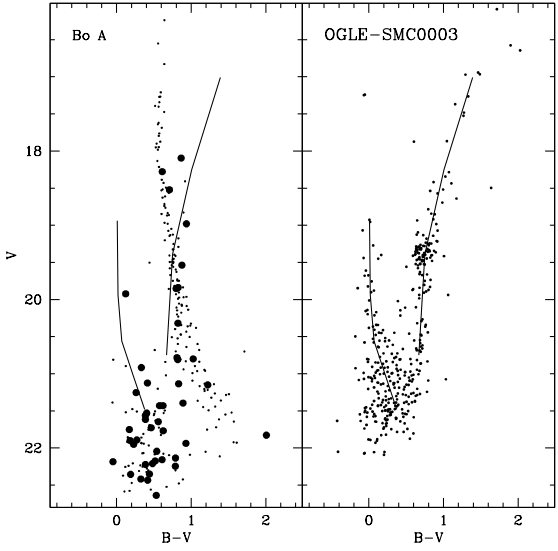}
   \caption{Left panel: CMD of all the stars within $r=1\arcmin$  from
   the center of Bo A (tiny dots). Stars with $r\le 15\arcsec$ are
   highlighted (large filled circles). The thin sequence of stars going
   from $V\sim 17$, $B-V\simeq 0.8$ to $V\sim 22$, $B-V\simeq 1.6$ is
   the MS of 47 Tuc. The ridge lines of the upper MS and of the RGB of
   OGLE-SMC0003 (\cite{pietr}) are overplotted on both panels. Right
   panel: CMD of the Small Magellanic Cloud cluster OGLE-SMC0003 from
   \cite{pietr}. }
   \label{Fig2}
    \end{figure}

The resulting Color Magnitude Diagram (CMD) of Bo~A is shown in
Fig.~\ref{Fig2} (left panel), where we plotted only stars within
$1\arcmin$ from the center of Bo~A (small dots). Stars lying within the
apparent radius of Bo~A ($r<15\arcsec$) are highlighted as large filled
circles. The CMD is dominated by the Main Sequence (MS) of 47~Tuc
(going approximately from $V\sim 17$, $B-V\simeq 0.8$ to $V\sim 22$,
$B-V\simeq 1.6$), while most of the stars within the apparent body of
Bo~A appear quite faint ($V>20.0$) and blue ($B-V<0.8$). 

\subsection{Bo~A as an overdensity of blue stars}

\begin{figure}
   \centering
   \includegraphics[width=8cm]{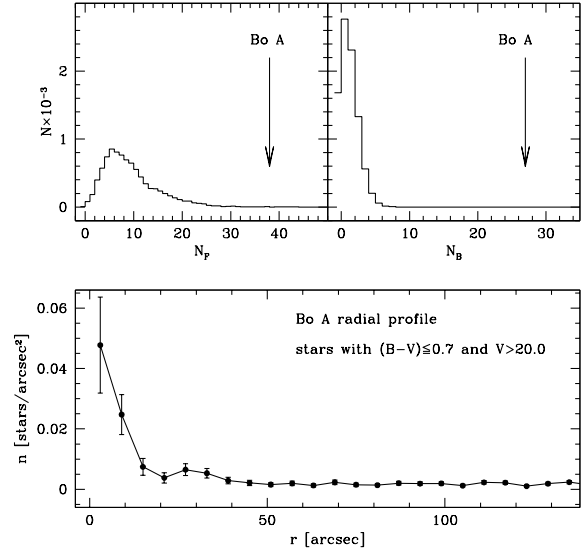}
   \caption{Upper Left Panel: Stars counted in 10\,000 randomly
   placed circles of radius $15\arcsec$ in the field around Bo~A, having
   $V>20.0$. The number of stars effectively counted in the circle
   centered on Bo~A ($N_F=38$) is marked by the arrow. Upper Right: Same
   as above, but only for blue stars, having $V>20.0$ and $B-V\le0.7$.
   The number counted in the circle around Bo~A ($N_B=27$) is marked by
   the arrow. Lower Panel: radial density profile of Bo A, obtained
   using faint blue stars ($V>20.0$ and $B-V\le0.7$) and $6\arcsec$
   bins. The profile flattens outside $r\sim 20\arcsec$ and remains flat
   over the whole chip 4 field (not shown here for reasons of scale).}
   \label{Fig3}%
   \end{figure}

We now use the CMD of Fig.~\ref{Fig2} to assess the statistical
significance of Bo~A as an overdensity of resolved stars with respect to
the underlying field of 47~Tuc. There are 38 stars with $V>20.0$ lying
within $r\le 15\arcsec$ from the center of Bo~A. To study the
probability of a chance occurrence of such a clustering, we counted the
total number of faint ($V>20.0$) stars ($N_F$) in 10\,000 circles,
having $r=15\arcsec$, randomly placed on chip 4. We added the further
condition that their centers avoid a region of radius $r=2\arcmin$ from
the center of Bo~A\footnote{We recall here that the field covered by
chip 4 is $\simeq 8.1\arcmin \times 16.2\arcmin$ and that it samples the
outer halo of 47~Tuc, hence the crowding is moderate all over the chip,
and finally Bo~A is not located where the density of stars of 47~Tuc is
the highest.}. The average over the 10\,000 random extractions is
$<N_{F}>=9.7 \pm 3.1$ (Poisson statistics), therefore $N_F=38$ measured
in Bo~A is significantly different at the $5.5 \sigma$ level. This is
graphically illustrated in the top left panel of Fig.~\ref{Fig3}, where
it can also be appreciated that $N_F$ values equal to or higher than 38
happen in 0.1\% of the cases only. Note that the adopted Poisson
statistics provides an error which is fully compatible with the observed
distribution.

%
%

Moreover, as we already noted, Bo~A appears bluer than the surrounding
field, with 27 out of 38 faint stars bluer than $B-V\le0.7$. We
therefore repeated the experiment, this time counting the number of
faint ($V>20.0$) and blue ($B-V\le0.7$) stars ($N_B$) in 10\,000
circles of $r=15\arcsec$, randomly placed on chip 4 as above. The
average  over 10\,000 random extractions is $<N_{B}>=1.7 \pm 1.3$
(Poisson statistics), where the $N_B=27$ measured for Bo~A is
significantly different, at the $5 \sigma$ level. The result is
illustrated in the top right panel of Fig.~\ref{Fig3}, where it can be
noted that the maximum value obtained for $N_B$ is only 8. 

Finally, if we consider the ratio of faint blue stars to faint red
stars $N_B/N_R$ (where faint red stars are those having $B-V>0.7$ and
$V>20.0$) we obtain $<N_{B}/N_{R}>=0.1 \pm 1.3$ over 10\,000 random
extractions, to compare with the observed value of $N_B/N_R=2.45 \pm 0.88$ for
Bo~A. Therefore, we can conclude not only that Bo~A is a statistically
significant aggregation of resolved stars in the field of 47~Tuc, but
also that its stellar population differs from that encountered anywhere
else in chip~4, being significantly bluer.

   \begin{figure}
   \centering
   \includegraphics[width=8cm]{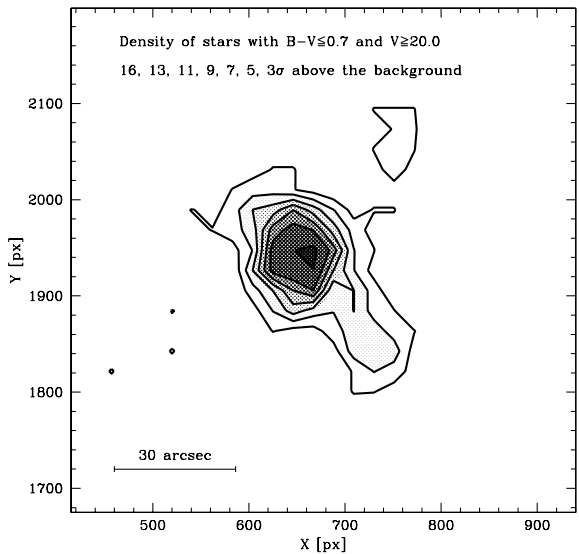}
   \caption{$2' \times 2'$ cutout of the density map of faint blue
   stars ($V>20.0$ and $B-V<0.7$) on chip 4. The map is computed by
   counting stars in circles of $r=10\arcsec$, placed on a grid spaced
   by $5\arcsec$. The average density (computed over the whole chip)
   has been subtracted and the remaining over-density has been divided
   by the standard deviation of the background, hence it is expressed
   in units of $\sigma$. Only around Bo~A the density of stars reaches
   a level of $5 \sigma$ above the background, while the peak density
   is more than $16 \sigma$.}
              \label{Fig4}%
    \end{figure}
%
%

\subsection{Shape, Size and Density Profile}

The density contours map of blue stars has been computed on circles of
radius $r=10\arcsec$, placed on a grid spaced by $5\arcsec$. The only
structure emerging above the $5 \sigma$ level on the background is Bo~A,
shown in Fig.~\ref{Fig4}, and the peak density is more than $16 \sigma$
higher than the background. A radial density profile has also been
computed in $6 \arcsec$ bins, using the faint blue stars as tracers of
Bo~A (Fig.~\ref{Fig3}, lower panel). It can be readily appreciated that
the density of these stars becomes essentially flat outside $\sim
20\arcsec$ from the center of the system and remains flat out to the
edges of the considered field.

Summarizing, Bo~A appears as a quite concentrated, approximately
circular overdensity, with a characteristic apparent diameter of
$D\simeq 30\arcsec$ (see Figs.~\ref{Fig1} and \ref{Fig3}). We note that
according to BD00 the typical diameters of confirmed SMC clusters range
from $\ge 12\arcsec$ to $\sim 4\arcmin$, where 25\% of the whole sample
has $D\le 30\arcsec$. Hence, the characteristic size of Bo~A is fully
compatible with the hypothesis that it is a cluster of the SMC galaxy.
In this hypothesis, and assuming a distance modulus of $(m-M)_0=18.82$
(\cite{mateo}) for the SMC, the linear size of Bo~A would therefore be
of approximately 8.4~pc.

\subsection{Stellar content}

Now that we have established the statistical significance of Bo~A as an
aggregate of resolved stars, we can use the CDM of Fig.~\ref{Fig2} to
study its stellar content. The ridge lines of the upper MS (the blue
plume at $B-V<0.6$) and of the Red Giant Branch (RGB, going from $V\sim
21$ and $B-V\sim 0.8$ to $V\sim 17$ and $B-V\sim 1.4$) of OGLE-SMC0003
are overplotted to both CMDs in Fig.~\ref{Fig2}. We can conclude from
this comparison that the CMD of Bo~A is fully compatible with that of a
sparse cluster belonging to the SMC. In this case the plume of blue
stars, lying around the extrapolation of the template MS ridge line at
faint magnitudes, should be interpreted as the upper MS of the
cluster, while most of the redder stars may belong to its RGB. 

The lower luminosity of the bright end of the MS could imply that
the age of Bo~A is larger than that of OGLE-SMC0003\footnote{In the
hypothesis that the two clusters have approximately the same
disctance.}, although small statistics prevent us from reaching a firm
conclusion on this point. The presence of a fully developed RGB and of a
well defined Red Clump would imply that OGLE-SMC0003 is older than 2-3
Gyr, which can provide a tentative lower limit to the age of Bo~A as
well. However, it has to be noted that the putative RGB stars of Bo A
are still fully compatible in color, magnitude and number with being MS
stars of 47~Tuc. There are in fact 8 stars within $15\arcsec$ from the
center of Bo~A with $0.4<B-V<1.1$ and $V<20.5$. In a set of 10\,000
random extractions as those described above, 8 or more  stars with these
characteristics are found in 20\% of the cases, hence there is no
significant overdensity of RGB stars around Bo~A. A fainter MS with
respect to OGLE-SMC0003, coupled with the lack of a corresponding RGB
would point against the hypothesis that Bo~A is an ordinary SMC cluster
and may instead suggest that is a much farther system (this possibility
is briefly considered in Sect.~4, below). In any case, low number
statistics prevent us from making any firmer statement, at the present
stage.

It is clear that a better characterization of the stellar
content of Bo~A requires a much deeper and higher resolution
photometry. We therefore searched various scientific archives for
better images of a field including Bo~A, but the only promising dataset
(a set of $t_{exp}=900$ s images taken with the ESO-NTT) resulted
useless because of the very bad seeing under which they were acquired
($FWHM\ge 2.5\arcsec$). Hence, dedicated observations are needed for a
deeper insight.

\section{Conclusions}

We have identified a previously unknown stellar system (Bo~A) in the 
background of the galactic globular cluster 47~Tucanae, located  $\sim
15\arcmin$ North-West of its center. The system, which we call Bo~A, is
partially resolved into stars and is clearly detected as a 
statistically significant overdensity of faint and blue stars in the
considered $\simeq 8.1\arcmin \times 16.2\arcmin$ field (chip 4),
located in the outskirts of 47~Tuc. The appearance, the
characteristic size, the CMD and the position in the sky of Bo~A
indicate that it is most probably a stellar cluster belonging to the
Small Magellanic Cloud. The proximity of Bo~A to 47~Tuc is the most
likely explanation to why the system has not been discovered before. 

While the identification of Bo~A as a SMC cluster is most likely,
the available data (and, in particular, the shallow CMD) leave formally
open (at least) another interpretation (see Sect.~3.4, above).  A dwarf
irregular galaxy located at $\sim 1$ Mpc or more from us (like, for
instance, Sag DIG \cite{lee}) may appear similar to Bo~A if seen in the
background of 47~Tuc, whose diffuse brightness may hide the unresolved
body of the galaxy. Its CMD would appear very similar to that of Bo~A,
once the same limiting magnitude ($V\sim 22.5$) is attained (see, Fig.~2
of \cite{lee}). Hence, accurate photometry down to $V\sim 24$ is
required to definitely rule out this less likely, but still
viable, hypothesis.

\begin{acknowledgements}

Financial support to this research has been provided by the Agenzia
Spaziale Italiana (ASI) and the Ministero dell'Istruzione,
dell'Universit\`a e della Ricerca (MIUR). We are grateful to an
anonymous Referee for her/his detailed and useful suggestions.

\end{acknowledgements}

\end{document}